\documentclass[
]{ceurart}

\usepackage{listings}
\usepackage{tikz}
\usetikzlibrary{positioning, arrows.meta}

\begin{document}

\copyrightyear{2026}
\copyrightclause{Copyright for this paper by its authors.
  Use permitted under Creative Commons License Attribution 4.0
  International (CC BY 4.0).}

\conference{ISWC 2026: 25th International Semantic Web Conference, Poster and Demos, October 25–29, 2026, Bari, Italy}

\title{RepuLink: A Linked Data Platform for Accountable Trust}
\tnotemark[1]
\tnotetext[1]{Demo paper accompanying~\cite{repulink2026}. Source code and the RepuLink ontology are available at the project repository (\url{https://github.com/hiuiwb/RepuLink-Tool}).}

\author[1]{Wenbo Wu}[%
email=wenbo.wu@soton.ac.uk,
]
\cormark[1]
\address[1]{University of Southampton, Southampton, United Kingdom}

\author[1]{George Konstantinidis}[%
email=g.konstantinidis@soton.ac.uk
]

\cortext[1]{Corresponding author.}

\begin{abstract}
Trust and reputation systems underpin reliable interactions in large, distributed networks. However, conventional models typically propagate trust only forward, offering no accountability for endorsers regarding whom they vouch for, and leaving newly joined nodes without a meaningful initial reputation. RepuLink addresses these limitations by proposing a two-layer trust and reputation model that integrates direct interaction feedback with domain-specific endorsements. Crucially, it holds endorsers accountable via Backward Endorsement Penalty/Reward Propagation (BEPP/BERP).
This paper demonstrates \emph{RepuLink-Tool}, a deployable, full-stack reference implementation of this model. The application enables nodes to interact, rate, and endorse each other, while tracking reputation via a live dashboard and an interactive trust network graph. Furthermore, we introduce a new Linked Data layer built on top of the application. This layer features a lightweight OWL ontology encompassing nodes, interactions, ratings, endorsements, pairwise trust assessments, and computed reputation scores annotated with PROV-O provenance. It also provides an on-the-fly RDF projection of each user's trust network in multiple serialisations, alongside a scoped SPARQL endpoint that nodes can query live against their own data.
\end{abstract}

\begin{keywords}
  trust and reputation \sep
  accountability \sep
  backward propagation \sep
  linked data \sep
  RDF \sep
  SPARQL \sep
  provenance
\end{keywords}

\maketitle

\section{Introduction}

Trust is the foundation of reliable interactions in distributed networks~\cite{wu2025trust}, underpinning applications such as data markets~\cite{ma2024model} and other settings where participants must decide whom to interact with based on limited direct experience. Trust and reputation management systems~\cite{josang2007survey} address this by aggregating feedback into a per-node score, often by \emph{propagating} trust through the network in the way PageRank~\cite{page1999pagerank} and EigenTrust~\cite{kamvar2003eigentrust} propagate importance. Two limitations recur across this line of work: newly joined or sparsely connected nodes cannot be reliably evaluated (the \emph{cold-start} problem) since there is no interaction history to draw upon; and propagation runs strictly forward, meaning a node that endorses a bad node faces no consequences when that node subsequently misbehaves.

RepuLink~\cite{repulink2026} addresses both limitations through a two-layer model comprising an interaction-feedback layer (representing pairwise local trust derived from direct ratings) and a domain-specific endorsement layer. The endorsement layer affords newly joined nodes an explainable initial reputation proportional to their endorsers' own reputation. These layers are combined via forward propagation into a single, network-wide reputation score for each node. The model incorporates two backward-propagation mechanisms (BEPP and BERP) that recursively penalise or reward a node's endorsers based on that node's subsequent behaviour.

Trust and reputation have a long history as first-class concerns on the Semantic Web, extending beyond the general trust-management literature upon which RepuLink~\cite{repulink2026} draws. Richardson et al.~\cite{richardson2003trust} proposed one of the earliest trust-propagation frameworks explicitly for the Semantic Web, evaluated on the same Epinions social network. Golbeck et al.~\cite{golbeck2003} introduced FOAF-based trust networks and metrics for propagating trust ratings along a social graph, and Artz and Gil~\cite{artzgil2007} survey the broader landscape of trust models proposed for Semantic Web applications. Closest in spirit to our provenance angle, Ceolin et al.~\cite{ceolin2012} combine user reputation with PROV-based provenance analysis to assess the trustworthiness of Web content. 
However, none of the prior work formalises \emph{backward-propagated accountability}, which is the process of recursively penalising or rewarding endorsers for the behaviour of their endorsee, as a machine-readable concept within an ontology. Existing models merely treat trust and provenance as numerical inputs for a scoring algorithm. They fail to represent the accountability relationship itself as a transparent, auditable graph structure.

This paper demonstrates \emph{RepuLink-Tool}, a deployable web application implementing the full model end-to-end, alongside a novel Linked Data layer that forms the core focus of this demonstration. The trust and endorsement graph inherent to the model maps naturally onto RDF: nodes, interactions, ratings, endorsements, and reputation scores are modelled as resources with typed relationships, rather than mere rows within a relational schema. By projecting the existing, pre-computed data through a lightweight ontology, we expose it as RDF across multiple serialisations and allow nodes to query it directly using SPARQL. This approach transforms an accountability mechanism that is typically opaque (often reduced to a single numeric score per user) into a fully explorable and auditable graph.

RepuLink-Tool's Semantic Layer builds directly on established Semantic Web standards---RDF~\cite{rdf11} and OWL~\cite{owl2} for the ontology and data model, FOAF~\cite{foaf} for agent identity, PROV-O~\cite{provo} for reputation provenance, and SPARQL~\cite{sparql11} for querying. It follows the established pattern of exposing an existing relational system as Linked Data through an on-demand mapping, rather than migrating its storage: the same principle formalised generically for relational databases by R2RML~\cite{r2rml}. We specialise this pattern to the trust and reputation domain, treating backward-propagated accountability (BEPP/BERP) and PROV-O-annotated reputation scores as first-class ontology concepts, rather than viewing RDF exposure as a generic schema-to-triples mapping exercise.

\section{System Architecture}
\label{sec:architecture}

\begin{figure}[h]
  \centering
  \begin{tikzpicture}[
      node distance=10mm,
      box/.style={draw, rounded corners, minimum width=32mm, minimum height=8mm, align=center, font=\small},
      newbox/.style={box, dashed, thick},
      ext/.style={box, fill=gray!10},
      arr/.style={-{Latex[length=2mm]}}
    ]
    \node[box] (fe) {React Frontend};
    \node[box, below=of fe] (be) {FastAPI Backend};
    \node[box, below left=10mm and -4mm of be] (db) {PostgreSQL};
    \node[box, below right=10mm and -4mm of be] (rep) {Reputation Engine\\(BEPP/BERP)};
    \node[newbox, right=14mm of be] (sem) {Semantic Layer\\RDF / SPARQL\\(this demo)};
    \node[ext, above right=10mm and 6mm of be] (kc) {Keycloak +\\User Mgmt. Service};

    \draw[arr] (fe) -- (be);
    \draw[arr] (be) -- (db);
    \draw[arr] (be) -- (rep);
    \draw[arr] (rep) -- (db);
    \draw[arr] (be) -- (sem);
    \draw[arr] (sem) -- (db);
    \draw[arr] (be) -- (kc);
  \end{tikzpicture}
  \caption{RepuLink-Tool architecture. The Semantic Layer (dashed) is a
    read-only RDF/SPARQL projection added on top of the existing
    relational data and reputation engine}
  \label{fig:architecture}
\end{figure}
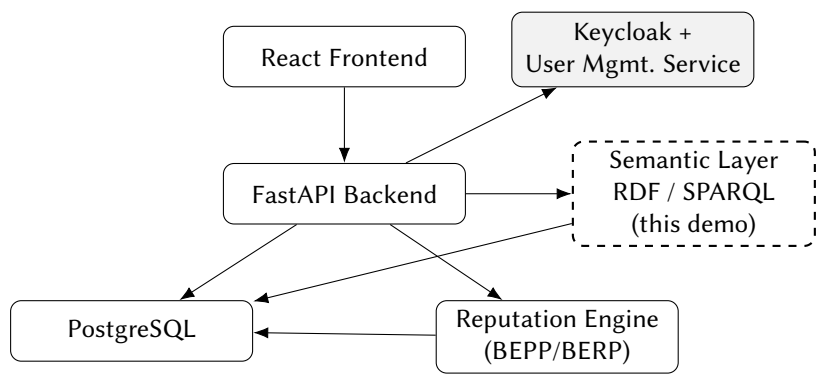

RepuLink-Tool (Figure~\ref{fig:architecture}) comprises a FastAPI and PostgreSQL backend paired with a React frontend, whilst identity management is delegated to an external Keycloak realm. The reputation engine is a pure Python module that recomputes every node's reputation following each relevant event, subsequently appending the result to a per-node history table. The Semantic Layer reads the same relational data already utilised by the rest of the application and produces RDF on request. Crucially, it does not introduce a native triple store, nor does it alter how reputation is computed or stored.

We deliberately chose to project existing relational data into RDF on demand, rather than migrating storage to a native triple store or maintaining a synchronised copy of the data. This approach leaves the fully implemented and tested reputation engine and its underlying storage completely untouched, preventing consistency drift between dual data stores, and requiring no additional long-running services. Specifically, each SPARQL query (Section~\ref{sec:semantic}) builds an in-memory graph using \texttt{rdflib}, evaluates the query against it, and immediately discards it---a process strictly scoped to a single request and a single node's visible network. 

\section{RepuLink as Linked Data}
\label{sec:semantic}

\subsection{Ontology}

We define a lightweight OWL ontology (using the namespace prefix \texttt{rl:}) that encompasses the resources outlined in Table~\ref{tab:ontology}. This ontology explicitly reuses FOAF~\cite{foaf} for agent identity and PROV-O~\cite{provo} to attach provenance metadata to the computed reputation scores. Within the RDF representation, a node from the underlying model (Section~\ref{sec:model}) is instantiated as an \texttt{rl:Agent} individual.

\begin{table}[h]
  \caption{RepuLink ontology classes.}
  \label{tab:ontology}
  \small
  \begin{tabular}{ll}
    \toprule
    Class & Key properties \\
    \midrule
    \texttt{rl:Agent} (\(\sqsubseteq\) \texttt{foaf:Agent}) & \texttt{foaf:name}, \texttt{foaf:mbox} \\
    \texttt{rl:Interaction} & \texttt{hasInitiator}, \texttt{hasTarget}, \texttt{hasStatus} \\
    \texttt{rl:Rating} & \texttt{ratingValue}, \texttt{ratedBy}, \texttt{ratedIn} \\
    \texttt{rl:Endorsement} & \texttt{endorser}, \texttt{endorsed}, \texttt{confidence} \\
    \texttt{rl:TrustAssessment} & \texttt{assessedBy}, \texttt{assessedFor}, \texttt{trustValue} \\
    \texttt{rl:ReputationScore} & \texttt{scoreOf}, \texttt{scoreValue}, \texttt{prov:wasGeneratedBy} \\
    \bottomrule
  \end{tabular}
\end{table}

The \texttt{rl:Interaction.hasStatus} property ranges over a concise controlled vocabulary (\texttt{rl:Pending}, \texttt{rl:Accepted}, \texttt{rl:Denied}) rather than relying on unconstrained string literals. Furthermore, every \texttt{rl:ReputationScore} individual is linked via \texttt{prov:wasGeneratedBy} to an activity individual representing the specific reputation computation that produced it. This deliberate, lightweight integration frames reputation as an \emph{auditable} computational process, rather than an opaque numerical value.

\subsection{Endpoints}

Four distinct endpoints expose this RDF projection. To strictly preserve privacy, all endpoints are scoped exclusively to the requesting node's visible network; no endpoint permits access to another node's private ratings:

\begin{itemize}
\item \texttt{GET /semantic/ontology} -- returns the ontology itself, serialised as Turtle.
\item \texttt{GET /semantic/network} -- returns the caller's own nodes, interactions, ratings, endorsements, trust assessments, and reputation scores, serialised as Turtle, JSON-LD, RDF/XML, or N-Triples (specified via the \texttt{?format=} parameter).
\item \texttt{GET /semantic/network/elements} -- returns the identical graph flattened into typed nodes and labelled property edges, utilised to render the interactive instance-graph view (Figure~\ref{fig:rdfgraph}).
\item \texttt{POST /semantic/sparql} -- executes a SPARQL query against the caller's own network graph (constructed dynamically per request) and returns standard SPARQL 1.1 JSON results~\cite{sparql11}.
\end{itemize}

\begin{lstlisting}[caption={Example query executed live during the demonstration: retrieving every node's reputation score, with names resolved via FOAF.}, label=lst:sparql]
PREFIX rl: <https://repulink.example/ontology#>
PREFIX foaf: <http://xmlns.com/foaf/0.1/>
SELECT ?name ?score WHERE {
  ?r a rl:ReputationScore ;
     rl:scoreOf ?agent ;
     rl:scoreValue ?score .
  ?agent foaf:name ?name .
} ORDER BY DESC(?score)
\end{lstlisting}

\section{System Demonstration}
\label{sec:demonstration}

We now recap the model, summarise the application it underpins, and walk through a live demonstration exercising all system components, including the newly introduced Linked Data layer.

\subsection{The RepuLink Model}
\label{sec:model}

We summarise the model only to the extent necessary to motivate the demonstration; comprehensive details, proofs, and evaluations are available in~\cite{repulink2026}. Each accepted interaction between two nodes may be rated on a scale of -5 to +5. From the cumulative positive and negative ratings between a pair of nodes, RepuLink derives a normalised \emph{local trust score}. Separately, nodes may \emph{endorse} one another with a confidence value, thereby forming a social graph independent of direct interaction history. A single \emph{reputation} score per node is computed by forward-propagating trust through both layers, projected onto the probability simplex. Crucially, a node's accumulated negative or positive feedback also adjusts the effective weight of \emph{their endorsers'} influence going forward (via BEPP/BERP). Consequently, reputation is not solely a function of a node's own behaviour, but also recursively depends upon the trustworthiness of those who vouched for them.

RepuLink-Tool recomputes this network-wide reputation following every rating or endorsement event, persists a per-node history, and surfaces both the pairwise local trust score and the global reputation score throughout the application (e.g., in search results, a per-node dashboard trend, and an interactive Trust Network graph).

\subsection{Application Overview}
\label{sec:overview}

\textbf{Dashboard.} A node's landing page displaying their current global reputation score, their rank amongst all registered nodes, and a line chart of their reputation history, which is recomputed and updated following every rating or endorsement event.

\noindent
\textbf{Interactions.} Nodes can search for one another by name, email, or exact identifier (including on-demand provisioning for a node known to the identity provider but not yet seen locally). They can send interaction requests, and accept or deny incoming ones. Once accepted, either party may rate the interaction (-5 to +5, with an optional comment). Each rating is explicitly labelled with its \emph{source} and \emph{target}, accommodating the fact that a single interaction can yield bilateral ratings.

\noindent
\textbf{Endorsements.} Nodes directly vouch for one another with a confidence value in \([0,1]\), independent of any prior interaction history. This constitutes the social endorsement layer through which the reputation model forward-propagates trust, and is the layer where BEPP/BERP enforces accountability for an endorsed node's subsequent behaviour.

\noindent
\textbf{Trust Network.} An interactive, per-node node-link visualisation (built using React Flow~\cite{reactflow}) mapping every node they have rated. Node colour encodes the local trust score, node size encodes interaction volume, and clicking a node reveals the comprehensive interaction and rating history with that specific node. Nodes are freely draggable for manual layout adjustment.

\noindent
\textbf{Linked Data.} The RDF projection and SPARQL endpoint introduced in Section~\ref{sec:semantic} function not only as backend services but also as a dedicated interface within the application. Accessible via the primary navigation, it features an RDF Instance Graph panel, a raw RDF viewer with a serialisation-format selector, and a SPARQL Playground with runnable example queries executed directly against the node's own live data.


\subsection{Live Walkthrough}
\label{sec:demo}

The live demonstration guides attendees through the system using two newly created accounts to reflect standard application usage. The walkthrough begins with two attendees signing up, completing an interaction, rating it, and mutually endorsing one another. Immediately, their respective dashboards update to visualise these events as new data points on their reputation history charts and adjust their global rankings. This immediate feedback demonstrates that the BEPP/BERP-adjusted reputation is computed live rather than batch-processed offline. Furthermore, because both attendees are brand-new nodes, this illustrates the cold-start mechanism (Section~\ref{sec:model}) in action, as each node's initial reputation instantly reflects the other's endorsement. Attendees can then explore this relationship via the interactive Trust Network graph, a draggable node-link visualisation where node colour signifies the local trust score, node size indicates interaction volume, and clicking a node reveals the comprehensive interaction history.

\begin{figure}[t]
  \centering
  \includegraphics[width=\linewidth]{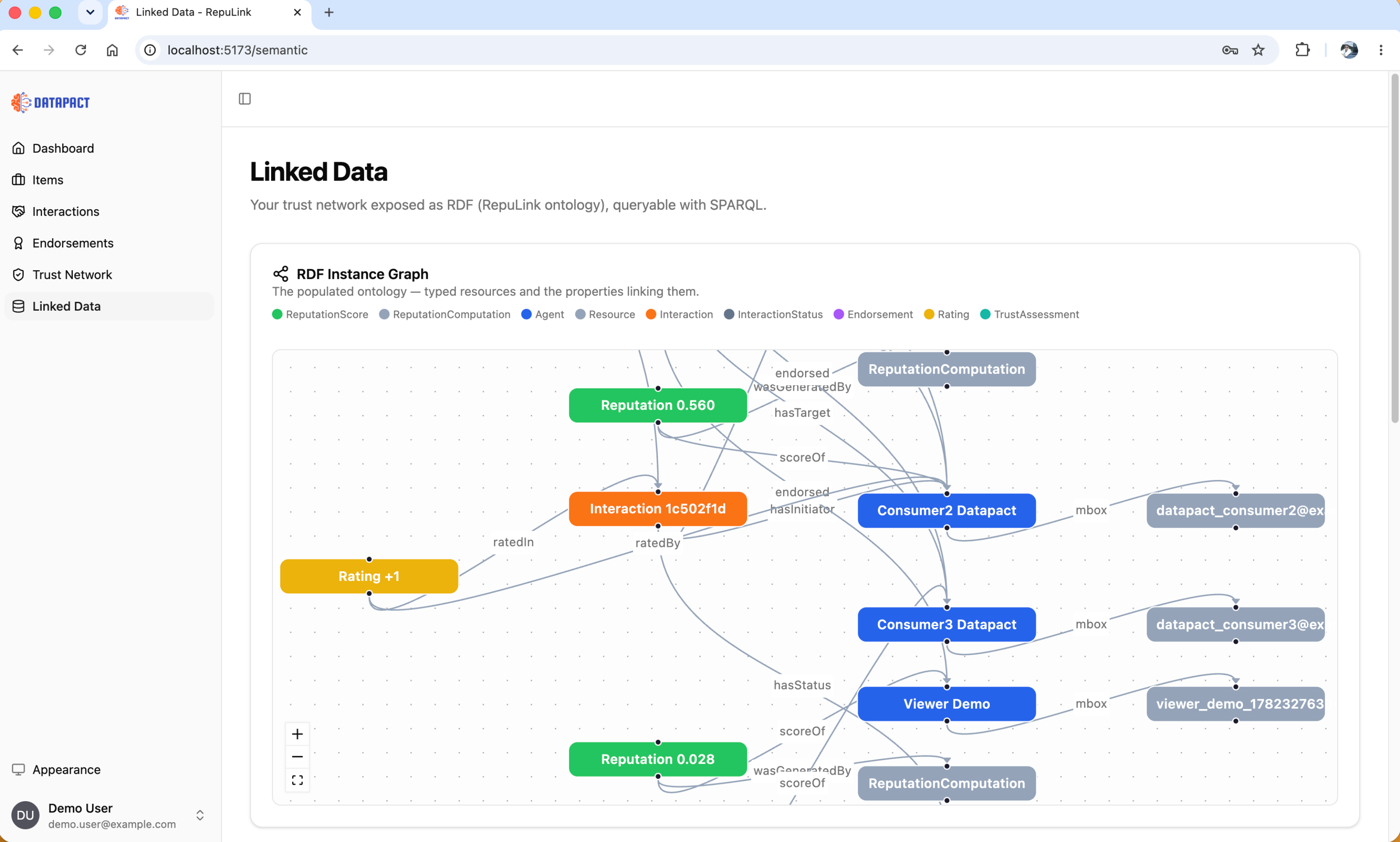}
  \caption{The Linked Data page's RDF Instance Graph panel, visualised live on a populated network.}
  \label{fig:rdfgraph}
\end{figure}

Transitioning to the semantic layer, the exact same data is subsequently rendered as a populated RDF Instance Graph (Figure~\ref{fig:rdfgraph}). In this view, typed resources---such as agents, interactions, ratings, and reputation scores---are connected by labelled object properties and automatically laid out, demonstrating that the application's underlying data is genuinely graph-shaped and ontology-conformant. To further interrogate their network, attendees can utilise the SPARQL Playground to execute and dynamically edit queries (e.g., Listing~\ref{lst:sparql}, or filters for highly trusted agents) directly against their own live data. Finally, the demonstration highlights the system's provenance capabilities by having attendees inspect an \texttt{rl:ReputationScore} individual's \texttt{prov:wasGeneratedBy} link. This crucial step illustrates how the BEPP/BERP accountability guarantee is rendered as a machine-checkable semantic property rather than merely asserted as an opaque numerical value.


\section{Conclusion and Future Work}
\label{sec:limitations}

We present a working demonstration of RepuLink's accountable, backward-propagated trust and reputation model. Furthermore, we demonstrated that its underlying interaction, endorsement, and reputation graph can be successfully exposed as genuine, queryable Linked Data. This is achieved via a lightweight ontology, multiple RDF serialisations, an interactive instance-graph view, and a live SPARQL endpoint, all implemented as a thin semantic layer atop an existing relational system, without necessitating alterations to its underlying computation or storage mechanisms. We posit that this design pattern---projecting an existing accountability mechanism into RDF rather than rebuilding it natively on a triple store---is broadly applicable to other trust and reputation systems seeking Semantic Web-native explainability.


\clearpage

\section*{Declaration of Use of Generative AI}
The author(s) acknowledge the use of Generative Artificial Intelligence (GenAI) in the preparation of this document. Specifically, GenAI was utilized as an advanced editing and text refinement tool. 
Its application was focused on enhancing the quality of pre-existing, human-authored text. This involved tasks such as: improving spelling and grammar, enhancing clarity and conciseness, and refining sentence structure for better engagement and flow.

\bibliography{repulink-ref}

@inproceedings{repulink2026,
  author    = {Wu, Wenbo and Konstantinidis, George},
  title     = {Enforcing Trust Accountability with Backward Propagation},
  booktitle = {Proceedings of the 32nd ACM SIGKDD Conference on Knowledge Discovery and Data Mining (KDD '26)},
  year      = {2026},
  note      = {To appear. Preprint: \url{https://arxiv.org/abs/2606.08851}}
}

@inproceedings{richardson2003trust,
  title={Trust management for the semantic web},
  author={Richardson, Matthew and Agrawal, Rakesh and Domingos, Pedro},
  booktitle={International semantic Web conference},
  pages={351--368},
  year={2003},
  organization={Springer},
  doi={https://doi.org/10.1007/978-3-540-39718-2_23}
}

@misc{wu2025trust,
      title={Trust and Reputation in Data Sharing: A Survey},
      author={Wenbo Wu and George Konstantinidis},
      year={2025},
      eprint={2508.14028},
      archivePrefix={arXiv},
      primaryClass={cs.SI},
      url={https://arxiv.org/abs/2508.14028},
}

@inproceedings{ma2024model,
  title={Model-based data markets: a multi-broker game theoretic approach},
  author={Ma, Yizhou and Jiang, Xikun and Wu, Evan W and Ib{\'a}{\~n}ez, Luis-Daniel and Shi, Jian},
  booktitle={2024 IEEE 23rd International Conference on Trust, Security and Privacy in Computing and Communications (TrustCom)},
  pages={2293--2301},
  year={2024},
  organization={IEEE}
}

@article{josang2007survey,
  title={A survey of trust and reputation systems for online service provision},
  author={J{\o}sang, Audun and Ismail, Roslan and Boyd, Colin},
  journal={Decision support systems},
  volume={43},
  number={2},
  pages={618--644},
  year={2007},
  publisher={Elsevier},
  doi = {https://doi.org/10.1016/j.dss.2005.05.019}
}

@techreport{page1999pagerank,
  title={The PageRank citation ranking: Bringing order to the web.},
  author={Page, Lawrence and Brin, Sergey and Motwani, Rajeev and Winograd, Terry},
  year={1999},
  institution={Stanford infolab}
}

@inproceedings{kamvar2003eigentrust,
  title={The eigentrust algorithm for reputation management in p2p networks},
  author={Kamvar, Sepandar D and Schlosser, Mario T and Garcia-Molina, Hector},
  booktitle={Proceedings of the 12th international conference on World Wide Web},
  pages={640--651},
  year={2003},
  doi={https://doi.org/10.1145/775152.775242}
}

@inproceedings{golbeck2003,
  author    = {Golbeck, Jennifer and Parsia, Bijan and Hendler, James},
  title     = {Trust Networks on the Semantic Web},
  booktitle = {Cooperative Information Agents VII (CIA 2003)},
  series    = {Lecture Notes in Computer Science},
  volume    = {2782},
  pages     = {238--249},
  publisher = {Springer},
  year      = {2003},
  doi       = {10.1007/978-3-540-45217-1_18}
}

@article{artzgil2007,
  author    = {Artz, Donovan and Gil, Yolanda},
  title     = {A Survey of Trust in Computer Science and the {S}emantic {W}eb},
  journal   = {Web Semantics: Science, Services and Agents on the World Wide Web},
  volume    = {5},
  number    = {2},
  pages     = {58--71},
  year      = {2007},
  doi       = {10.1016/j.websem.2007.03.002}
}

@inproceedings{ceolin2012,
  author    = {Ceolin, Davide and Groth, Paul and van Hage, Willem Robert and Nottamkandath, Archana and Fokkink, Wan},
  title     = {Trust Evaluation through User Reputation and Provenance Analysis},
  booktitle = {Proceedings of the 8th International Workshop on Uncertainty Reasoning for the Semantic Web (URSW 2012)},
  series    = {CEUR Workshop Proceedings},
  volume    = {900},
  pages     = {15--26},
  publisher = {CEUR-WS.org},
  year      = {2012}
}

@misc{foaf,
  author       = {Brickley, Dan and Miller, Libby},
  title        = {{FOAF} Vocabulary Specification 0.99},
  howpublished = {\url{http://xmlns.com/foaf/spec/}},
  year         = {2014}
}

@misc{provo,
  author       = {Lebo, Timothy and Sahoo, Satya and McGuinness, Deborah},
  title        = {{PROV-O}: The {PROV} Ontology},
  howpublished = {\url{https://www.w3.org/TR/prov-o/}},
  organization = {W3C Recommendation},
  year         = {2013}
}

@misc{rdf11,
  author       = {Cyganiak, Richard and Wood, David and Lanthaler, Markus},
  title        = {{RDF} 1.1 Concepts and Abstract Syntax},
  howpublished = {\url{https://www.w3.org/TR/rdf11-concepts/}},
  organization = {W3C Recommendation},
  year         = {2014}
}

@misc{sparql11,
  author       = {Harris, Steve and Seaborne, Andy},
  title        = {{SPARQL} 1.1 Query Language},
  howpublished = {\url{https://www.w3.org/TR/sparql11-query/}},
  organization = {W3C Recommendation},
  year         = {2013}
}

@misc{reactflow,
  key          = {React Flow},
  title        = {{React Flow}: Node-Based UIs in React},
  howpublished = {\url{https://reactflow.dev}},
  year         = {2024}
}

@misc{r2rml,
  author       = {Das, Souripriya and Sundara, Seema and Cyganiak, Richard},
  title        = {{R2RML}: {RDB} to {RDF} Mapping Language},
  howpublished = {\url{https://www.w3.org/TR/r2rml/}},
  organization = {W3C Recommendation},
  year         = {2012}
}

@misc{owl2,
  key          = {OWL 2},
  title        = {{OWL} 2 Web Ontology Language Document Overview},
  howpublished = {\url{https://www.w3.org/TR/owl2-overview/}},
  organization = {W3C Recommendation},
  year         = {2012}
}

\end{document}